\documentclass[11pt,a4paper]{article}
\pdfoutput=1

\usepackage{epsfig}
\usepackage{amsmath}
\usepackage{amssymb}
\usepackage{verbatim}
\usepackage{bm}
\usepackage{dsfont}
\usepackage[footnotesize]{caption}
\usepackage{subfigure}
\usepackage{cancel}
\usepackage{cite}
\usepackage{a4wide}
\usepackage{color}

\usepackage{multirow}

\def\Mp{{M_{\rm P}}}

\def\vp{\varphi}

\usepackage{geometry}
\begin{document}

\begin{minipage}{0.4\textwidth}
 \begin{flushleft}
DESY 14-099 \\
IPMU 14-0150\\
SISSA 39/2014/FISI \\
\end{flushleft}
\end{minipage} \hfill
\begin{minipage}{0.4\textwidth}
\begin{flushright}
 June 2014
\end{flushright}
 \end{minipage}

\vskip 1.5cm

\begin{center}
{\LARGE\bf The Chaotic Regime of D-Term Inflation}

\vskip 2cm

{\large W.~Buchm\"uller$^a$, V.~Domcke$^b$,  K.~Schmitz$^c$ }\\[3mm]
{\it{
a DESY, 22607 Hamburg, Germany \\
b SISSA/INFN, 34126 Triest, Italy \\
c Kavli IPMU (WPI),  Kashiwa 277-8583, Japan}
}
\end{center}

\vskip 1cm

\begin{abstract}
\noindent 
We consider D-term inflation for small couplings of the inflaton to matter fields. Standard hybrid inflation then ends at a
critical value of the inflaton field that exceeds the Planck mass. During the subsequent waterfall transition 
the inflaton continues its slow-roll motion, whereas the waterfall
field rapidly grows by quantum fluctuations. Beyond the decoherence time,
the waterfall field becomes classical and approaches a time-dependent minimum,
which is determined by
the value of the inflaton field and the self-interaction of the waterfall field.
During the final stage of inflation, the effective inflaton
potential is essentially quadratic,
which leads to the standard predictions of
chaotic inflation. The model illustrates how the decay of a false
vacuum of GUT-scale energy density can end in a period of `chaotic inflation'.
\end{abstract}

\thispagestyle{empty}

\newpage

\section{Introduction}

Chaotic inflation \cite{Linde:1983gd} and hybrid inflation \cite{Linde:1991km} are two
theoretically appealing frameworks for the description of the very
early universe. They are also the prototypes of the two different
versions of inflation: large-field and small-field inflation,
respectively. In their simplest form, both are not favoured by the
Planck data on the cosmic microwave background (CMB) \cite{Ade:2013uln}, but both are
consistent with the CMB data at the 2$\sigma$ level. The recently released BICEP2 results
\cite{Ade:2014xna} provide evidence for primordial gravitational waves and a
correspondingly large tensor-to-scalar ratio $r$, favouring
chaotic inflation and a GUT-scale energy density during
inflation. Although presently under intense scrutiny~\cite{Mortonson:2014bja, Flauger:2014qra}, these results indicate that current and upcoming experiments have now reached the sensitivity to probe inflation models which yield sizeable tensor modes.

In supersymmetric D-term hybrid inflation \cite{Binetruy:1996xj,Halyo:1996pp}, the inflaton field
can take large, transplanckian values if the coupling of the inflaton
to the waterfall fields is small. In principle, this offers the
possibility for a significant amplitude of primordial gravitational
waves. In standard hybrid inflation, however, the scalar-to-tensor
ratio is small since the inflaton potential is very flat.

In this paper we study the phase of tachyonic preheating \cite{Felder:2000hj} which
follows standard hybrid inflation when the inflaton passes the
critical value beyond which the waterfall field starts to grow by
quantum fluctuations. This is a complicated nonequilibrium process, which
has been studied numerically neglecting the Hubble  expansion
\cite{Felder:2000hj,Copeland:2002ku}. In the following, we discuss this process taking both the
Hubble expansion and the motion of the inflaton field into account.
The onset of tachyonic preheating can then be treated analytically
\cite{Asaka:2001ez}. Due to the rapid growth of quantum fluctuations, the waterfall
field reaches a classical regime within a few Hubble times, during which
backreactions can be neglected. In this way one obtains the initial conditions
for the subsequent classical evolution of the waterfall and inflaton
fields, where backreaction effects can be accounted for approximately by means of
the nonlinear classical field equations. As the waterfall field
approaches the global minimum of the scalar potential, its expectation value provides an
effective mass term for the inflaton and the system reaches a
regime of `chaotic inflation'. As we shall show, a successful
description of the observational data can then be obtained for typical
parameters of the model.

Inflation during the waterfall transition has been
studied before, however for a very different parameter regime. In
Refs.~\cite{Clesse:2010iz, Kodama:2011vs, Lyth:2012yp}, a small-field version of inflation has been
considered, where the inflaton field stays close to the critical value
and inflation proceeds essentially along a trajectory in the direction of the
waterfall field. This can account for a sufficient number of $e$-folds
and a red-tilted spectrum, yet the tensor-to-scalar ratio is very
small. In the parameter regime considered in this paper, a regime of
chaotic inflation emerges far away from a transplanckian critical point in a region of field space where inflation proceeds mainly in inflaton direction.

The paper is organized as follows. In Section~\ref{sec_2}, we recall essential
elements of D-term hybrid inflation. The main part of the paper is
Section~\ref{sec_3}, where the dynamics of tachyonic preheating are discussed
taking into account Hubble expansion and inflaton motion. Section~\ref{sec_4}
deals with the emerging regime of chaotic inflation and in Section~\ref{sec_5}
some aspects of cosmic string production are discussed. We conclude in Section~\ref{sec_6}.

\section{D-term hybrid inflation \label{sec_2}}

Let us start by briefly reviewing the setup of D-term hybrid inflation~\cite{Binetruy:1996xj, Halyo:1996pp}. Its main ingredients are a $U(1)$ gauge symmetry featuring a Fayet-Iliopoulos (FI) term as well as an $R$-invariant superpotential,
\begin{equation}
 W = \lambda \,\phi\, S_+ S_- \,,
\end{equation}
where the singlet $\phi$ contains the inflaton field and the waterfall
fields $S_\pm$ carry charge $\pm 1$ under the $U(1)$
symmetry. $\lambda$ is a dimensionless coupling constant, which
w.l.o.g.\ we can take to be positive. Furthermore, we work in a
supergravity framework employing a canonical K\"ahler potential, which respects a
shift symmetry, $R$ symmetry as well as a discrete $\mathbb{Z}_2$ symmetry \cite{Kawasaki:2000yn},
\begin{equation}
 K = \frac{1}{2}\left(\phi +  \bar\phi\right)^2 + S_+ \bar{S}_+ + S_- \bar{S}_- \,,
\end{equation}
so as to retain a sufficiently flat potential even for the large
values of the inflaton field expected from recent indications of a
sizable tensor-to-scalar ratio~\cite{Ade:2014xna}.\footnote{We thank
  Tsutomu Yanagida for pointing out this possibility of realizing
  D-term hybrid inflation with transplanckian field  excursions in
  supergravity. \smallskip} The introduction of an FI-term in
supergravity is a subtle
issue~\cite{Binetruy:2004hh,Komargodski:2010rb, Dienes:2009td}; it can
however be achieved employing strong dynamics in a field theory
setup~\cite{Domcke:2014zqa}.\footnote{For a recent discussion of field-dependent
  FI-terms for anomalous U(1) symmetries, cf.\ Ref.~\cite{Wieck:2014xxa}. \smallskip}

At large field values for the inflaton $\vp = \sqrt{2} \, \text{Im}(\phi)$, the fields $S_\pm$ are stabilized at zero and the tree-level potential for the inflaton is flat with the vacuum contribution being provided by the FI-term, $V_0 = \frac{1}{2} g^2 \xi^2$. Here, $g$ denotes the associated $U(1)$ gauge coupling constant. The Coleman-Weinberg one-loop potential, obtained by integrating out the heavy waterfall multiplets, lifts the flatness of the potential, so that above the critical point the total scalar potential is given by
\begin{equation}
 V(\varphi) = V_0 \left(1 + \frac{g^2}{16 \pi^2} \left( (x-1)^2 \ln(x-1) + (x + 1)^2 \ln(x+1) - 2 x^2 \ln x  - \ln 16 \right)\right) \,,
\label{eq_V1}
\end{equation}
with $x = \lambda^2 \vp^2/(2 g^2 \xi)$. The critical point $\vp_c$ where the $U(1)$ symmetry breaking field $s = \sqrt{2} |S_+|$ becomes tachyonic is determined by $x=1$,
\begin{equation}
 \vp_c = \frac{g}{\lambda} \,\sqrt{2 \xi} \,.
\label{eq_tauc}
\end{equation}
Below $\varphi_c$, the scalar potential depends non-trivially on both the inflaton and the waterfall field. Neglecting the higher-dimensional inflaton-waterfall couplings induced by supergravity, we have\footnote{Here and in the following, we work in units of the reduced Planck mass, $\Mp = (8 \pi G)^{-1/2} \equiv 1$.}
\begin{equation}
 V(\vp, s) = \frac{g^2}{8} (s^2 - 2 \xi)^2 + \frac{\lambda^2}{4} s^2\vp^2  + {\cal O}(s^4\vp^2) \,,
\label{eq_V2}
\end{equation}
until in the true vacuum at $\vp = 0$ and $s = s_0 \equiv \sqrt{2 \xi}$
supersymmetry is restored. The dynamics after the critical point are
usually assumed to proceed very fast, i.e.\ the waterfall field
undergoes a phase of tachyonic preheating, ensuring that $s$ rapidly
reaches its true minimum, whereas the homogeneous inflaton field 
quickly transitions to its true vacuum. 
In the next section, we show that for small values of the coupling
constant,\footnote{As the shift symmetry for Im($\phi$) is restored for $\lambda \rightarrow 0$, a small value of $\lambda$ is natural in the sense of 't Hooft.} $\lambda/g \ll 1$, and taking the Hubble expansion into account,
the picture is actually quite different: After passing through the critical point, tachyonic preheating indeed proceeds rapidly within a few $e$-folds, but the subsequent dynamics of the  homogeneous fields generate a large amount of $e$-folds, dramatically changing the predictions for the CMB observables.

\section{Tachyonic growth of quantum fluctuations \label{sec_3}}

At the critical point the waterfall field becomes tachyonic, which leads
to a rapid growth of the low-momentum ($k < k_*$) quantum fluctuations
and hence of the variance $\langle s^2(t)\rangle$. Neglecting the
Hubble rate, the variance becomes comparable to the global minimum  at
a spinodal time~\cite{Weinberg:1987vp} $t_{\rm sp} \sim
\mathcal{O}(1/m)$, where $-m^2$ is the tachyonic mass squared of the waterfall field in the quench approximation. The phase transition is found to be completed after a `single oscillation'~\cite{Felder:2000hj}. The root mean square value of the waterfall field can then be interpreted
as a homogeneous background field, $s(t) \simeq \langle s^2(t)\rangle^{1/2}$ within a patch of the size $\sim k_*^{-1}$.
During this phase transition, topological defects
are generically formed, which are separated by the coherence length $k_*^{-1}$.
The growth of the fluctuations is terminated by backreaction, i.e.\ by the self-interaction of the waterfall field, as the
 different modes scatter off each other. Tachyonic preheating is a complicated nonequilibrium process, which has been studied
numerically for hybrid inflation-type models, neglecting the Hubble expansion \cite{Felder:2000hj,Copeland:2002ku}. {For previous work on the effect of the Hubble expansion during preheating, cf.\ Refs.~\cite{Greene:1997ge, BasteroGil:1999fz}.}

In the case under consideration, the Hubble expansion and the (small) velocity $\dot \vp$ of the inflaton field are of crucial importance. In particular, tachyonic preheating ends at a `local spinodal time' $t_{\rm sp}^{\rm loc} \sim \mathcal{O}(1/H) $, when the quantum fluctuations become comparable with the instantaneous,
inflaton-dependent minimum.
Near the critical point
$\vp_c$, the inflaton motion is approximately linear in time. For this
case, the onset of tachyonic preheating has been studied analytically 
\cite{Asaka:2001ez}. It has been shown that quantum fluctuations grow
with time faster than exponentially and that the phase transition is completed
within a few Hubble times, if backreaction effects are small. In the case
of D-term inflation, the strength of the backreaction is given by the gauge
coupling, and hence strong. In the following we will therefore use
the method of Ref.~\cite{Asaka:2001ez} to compute the growth of the
waterfall field up to the decoherence time $t_{\rm dec}$, where the
field becomes classical and where the backreaction is still small.
For later times we shall take the backreaction approximately into
account by means of the nonlinear classical field equations. 
 
We are interested in a parameter regime where the dynamics of the
homogeneous inflaton field is slow compared to tachyonic preheating,
i.e.\ where the velocity $\dot \vp_c$ of the inflaton field when crossing
the critical point is small and thus the quench approximation is
inapplicable. In particular, we will find $\dot \vp_c \sim H^2$ with
$H$ denoting the Hubble parameter. Hence, contrary to the situation
in Ref.~\cite{Copeland:2002ku}, the Hubble expansion affects the dynamics
during preheating and cannot be neglected. 

Close to the critical point ($t_c = 0$), the potential for the waterfall field can be expressed as, cf.\ Eq.~\eqref{eq_V2},
\begin{equation}
 V(s; t) \simeq \frac{1}{2} g^2 \xi^2 - \frac{1}{2} D^3  t \, s^2 +
 {\cal O}( t^2, s^4) \ ,
\end{equation}
where $D^3 \equiv \sqrt{2 \xi} g \lambda |\dot \vp_c| $ and we have used
\begin{equation}
 \vp(t) \simeq \vp_c + \dot \vp_c  t \,.
 \label{eq_tau}
\end{equation}
The inflaton velocity is obtained from the slow-roll equation of motion for the inflaton in the scalar potential~\eqref{eq_V1},
\begin{equation}
 \dot \vp_c = - \frac{\partial_\vp V}{3 H} \bigg|_{\vp_c} = - \frac{g^{2} \lambda \ln 2}{4 \sqrt{3} \pi^2} \sqrt{\xi} \,.
\end{equation}
For a typical parameter choice, which will yield the correct amplitude of the primordial power spectrum, $g^2 = 1/2$, $\lambda = 5 \times 10^{-4}$ and $\sqrt{\xi} = 2.8 \times 10^{16}$~GeV, this implies
\begin{equation}
H_c \equiv H(\vp_c) = 9.1 \times 10^{13}~\text{GeV}\,, \quad  \dot \vp_c \simeq -21 \,H_c^2\,, \quad  D \simeq 1.5 \, H_c \,.
\end{equation}
Note that within a Hubble time the inflaton field changes only by
$\dot\varphi_c/H_c \sim 10^{-4} \varphi_c$.

{To study the growth of the quantum fluctuations of the waterfall field around the critical point, we decompose the waterfall field into its momentum eigenfunctions,}
\begin{equation}
 {s(t, \boldsymbol{x}) = e^{-\frac{3}{2} H t} \int \frac{d^3k}{(2 \pi)^{3/2}} \left( a_s(\boldsymbol{k}) s_k(t) e^{i \boldsymbol{k x}} +  a_s^\dagger(\boldsymbol{k}) s_k^*(t) e^{- i \boldsymbol{k x}} \right)\,.}
\end{equation}
{Here $a_s(\boldsymbol{k})$ and $a^\dagger_s(\boldsymbol{k})$ denote the annihilation and creation operators, $s_k(t)$ is the amplitude of the mode with fixed comoving momentum $k$. The time dependence of $s_k(t)$}
is determined in linear
approximation by the mode equation \cite{Asaka:2001ez},
\begin{align}
\ddot{s_k} + \left(k^2 e^{-2Ht} - \frac{9}{4} H^2\ - D^3 t\right)s_k =
0 \ .
\end{align}
There are two regions of momenta, which are separated by the boundary condition
$k = k_b(t)$, defined by
\begin{align}
\left.\left(k^2 e^{-2Ht} - \frac{9}{4} H^2\ - D^3
    t\right)\right|_{k=k_b(t)} = 0 \ .
\end{align}
Modes with $k > k_b(t)$ oscillate in time whereas modes with $k <
k_b(t)$ show tachyonic growth. For large times the latter are given by the
Airy functions Ai and Bi,
\begin{equation}
 s_k(t) \simeq i \sqrt{\frac{\pi}{2 D}} \, \text{Ai}(D t) + \sqrt{\frac{\pi}{2 D}} \, \text{Bi}(D t)  \,.
\end{equation}
Oscillating modes with $k > k_b(t)$ are given by Hankel functions. 

The growth of the waterfall field after the critical point due to
quantum fluctuations is given by the variance
\begin{align}
\langle s^2(t) \rangle = \langle s^2(t) \rangle_{\rm us} - \langle
s^2(0) \rangle_{\rm us} \ ,
\end{align}
where the index `us' marks the unsubstracted quantities before
renormalization,\footnote{Calculating this quantity numerically we
  introduce a comoving IR-cutoff at $k = H_c$ and a physical
  UV-cutoff at $k/a = 10 H_c$, cf.\ also Refs.~\cite{Seery:2010kh,
    Xue:2011hm}. As expected, our results are numerically independent of the choice of the UV-cutoff. The IR-cutoff corresponds to the $k$-resolution of the system, which is determined by $H_c$.
Since the dispersion is dominated by momenta $k > H_c$ (see comment below Eq.~\eqref{growth}) the choice of the IR-cutoff introduces a theoretical uncertainty of at most a few percent, which will not affect the main results of this paper.}
\begin{equation}
 \langle s^2(t) \rangle_{\rm us} = \int_0^{\infty} dk \,\frac{k^2}{2 \pi^2}\, e^{- 3 H_c t}\, | s_k(t)|^2 \,.
\end{equation}
Note that the variance $\langle s^2(t) \rangle$ is both ultraviolet and infrared
finite. Approximating $s_k$ above and below $k_b$ by Hankel and Airy
functions, respectively, one obtains for $Dt \gg 1$ the analytic
estimate \cite{abc}, 
\begin{align}\label{growth}
 \langle s^2(t) \rangle &\simeq 
\int_0^{k_b(t)} dk \,\frac{k^2}{2 \pi^2}\, e^{- 3 H_c t}\, | s_k(t)|^2  \nonumber\\
&\simeq \frac{3^{2/3}\Gamma^2\left(\frac{1}{3}\right)}{192\pi^3}D^2
(Dt)^{3/2}\exp{\left(\frac{4}{3}(Dt)^{3/2} + Ht\right)} \ .
\end{align}
For times $Dt > 1$ one has $D^3t > H_c^2$, and therefore $p_b(t) =
k_b(t) \exp{(- H_ct)} \simeq (D^3t)^{1/2} > H_c$. One easily
verifies that the integral \eqref{growth} is dominated by physical momenta
$\mathcal{O}(p_b)$, i.e. by modes inside the horizon $1/H_c$.

\begin{figure}
\begin{center} 
\includegraphics[width = 0.47\textwidth]{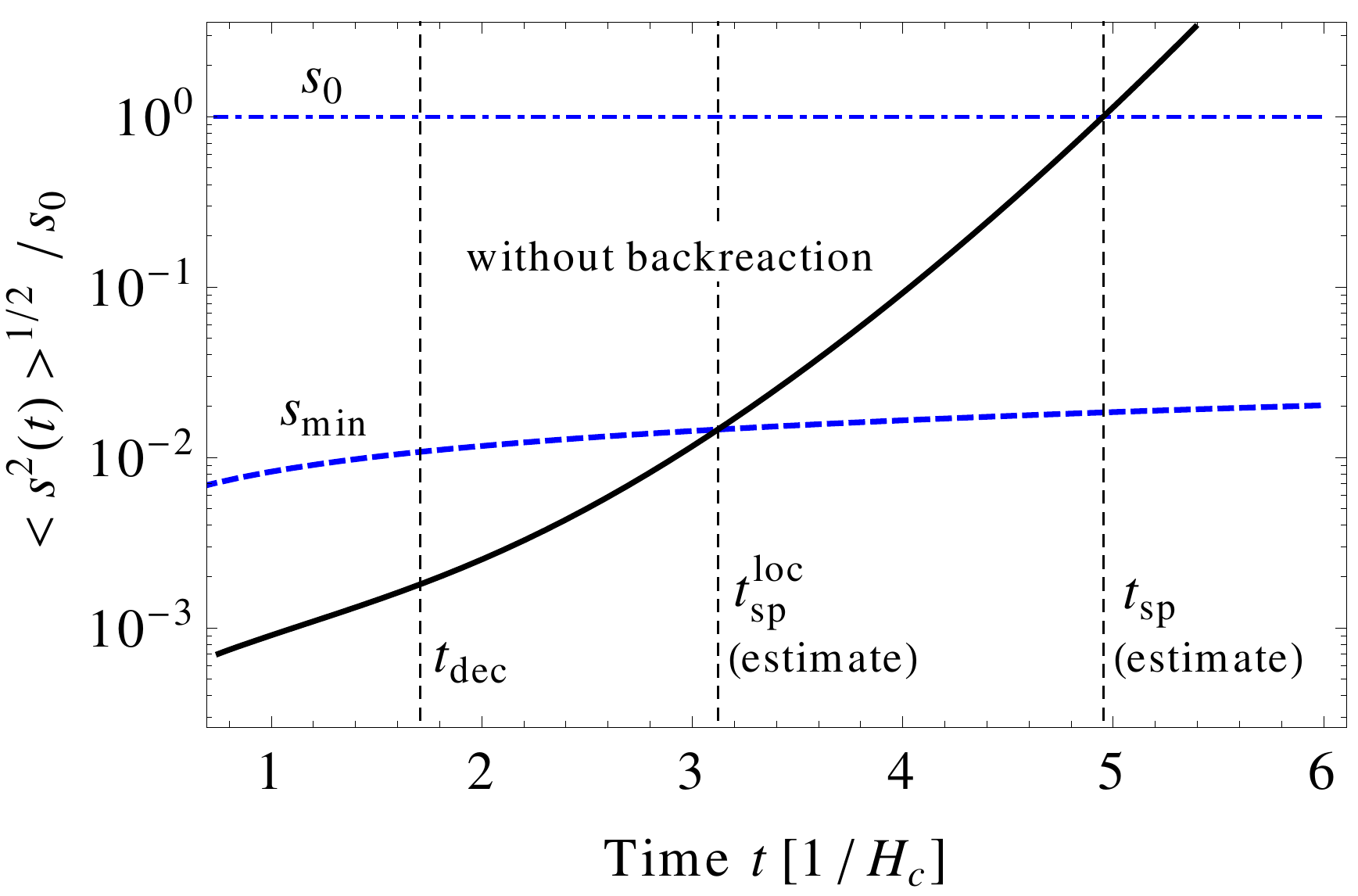}
\hspace{0.7cm}
\includegraphics[width = 0.47\textwidth]{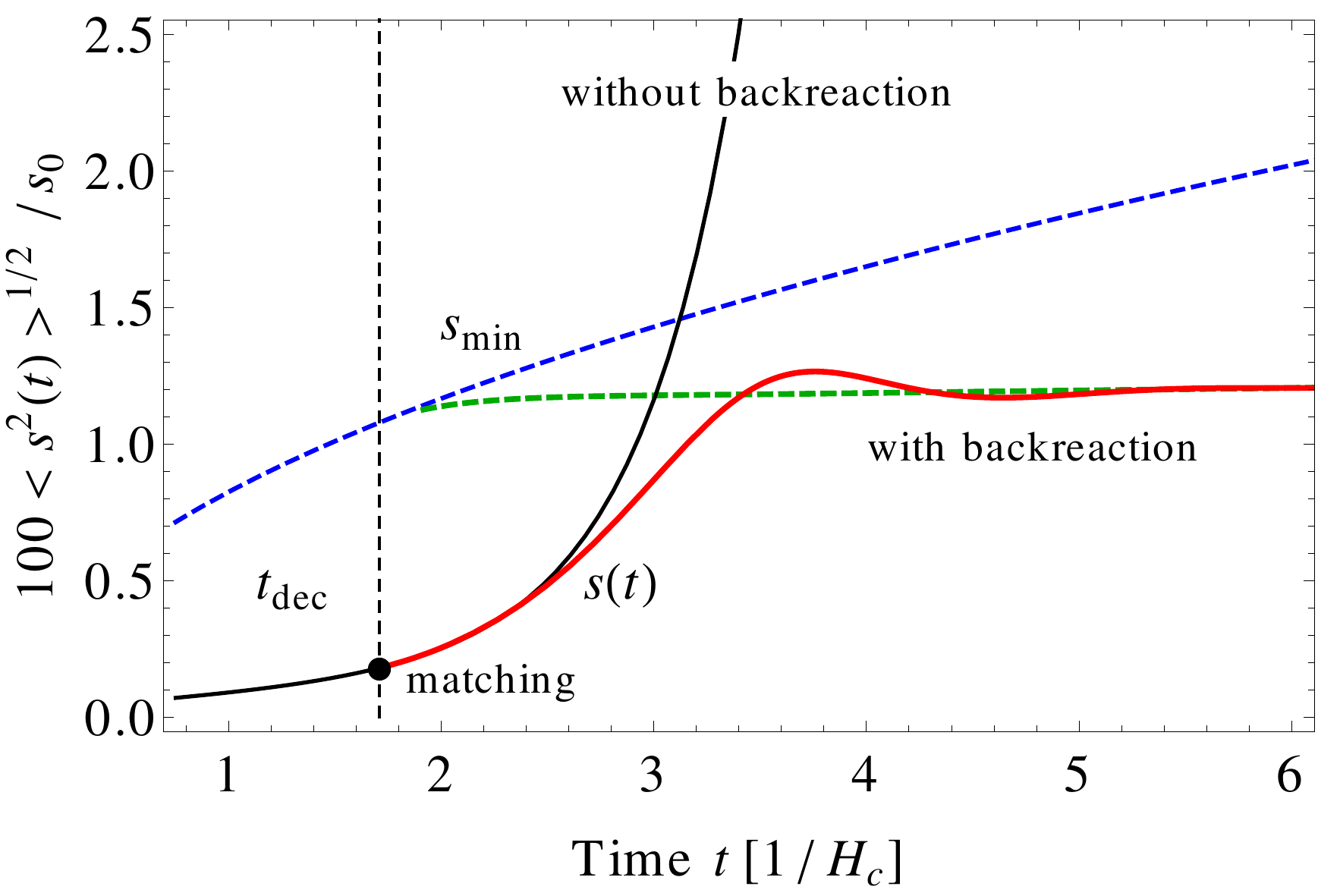}
\end{center} 
\caption{\textbf{Left (a):} Growth of the waterfall field during tachyonic preheating due to
 quantum fluctuations, comparison
   with inflaton-dependent minimum $s_{\textrm{min}}$ in the linear approximation~\eqref{eq_tau} (dashed thick blue) and global minimum $s_0$ (dashed
   thin blue) ignoring backreaction effects. Here, we determined the variance $\left<s^2(t)\right>^{1/2}$ numerically,
   i.e.\ by explicitly solving the mode equation for the waterfall field in the relevant $k$ range.
   \textbf{Right (b):} Classical evolution of the homogeneous waterfall field $s(t)$
   after the decoherence time (solid red), comparison with inflaton-dependent
   minimum including backreaction effects via classical field
   equations (dashed green). Note that the initial conditions for the evolution of $s(t)$
   in the right panel correspond to the numerical values of $\left<s^2(t)\right>^{1/2}$ and its derivative
   at $t = t_{\textrm{dec}}$ in the left panel.
   Here and in the following figures, the parameter values are: $g^2 = 1/2$,
   $\lambda = 5 \times 10^{-4}$ and $\sqrt{\xi} = 2.8 \times 10^{16}$~GeV. }
 \label{fig_tp}
\end{figure}

For small backreaction, an estimate of the spinodal time is
obtained from
\begin{equation}
 \langle s^2(t_{\rm sp}) \rangle \simeq 2 \,\xi\,.
\end{equation}
In the case of the above parameter example one finds, cf.\ Fig.~\ref{fig_tp}a, $t_{\rm sp}
\simeq 5.0/H_c$, i.e.\ the tachyonic growth would be completed within $N_{\rm
  tp} \simeq 5$ $e$-folds if the backreaction could be neglected. 
 This can be compared with the local spinodal time needed to
reach the inflaton-dependent minimum obtained from the potential \eqref{eq_V2},
\begin{equation}\label{s2min}
 s_\text{min}^2(\vp) = s_0^2 - \frac{\lambda^2}{g^2} \vp^2 \,, \quad s_0 = \sqrt{2 \xi} \,,
\end{equation}
which is sensitive to the self-interaction of the waterfall field
and which is relevant in our case, as discussed below.
From Fig.~\ref{fig_tp}a one reads off a value $t^{\rm loc}_{\rm sp} \simeq
3.1/H_c$. 

The tachyonic growth described above is very similar to the standard
picture of tachyonic preheating \cite{Felder:2000hj,gb1,gb2}. In the
simplest case where a constant tachyonic mass $m$ is turned on at  
$t=0$ and the Hubble parameter can be neglected, the variance is
also given by the integral \eqref{growth} with $k_b = m$ and $H_c =
0$. It is well known that in this case the false vacuum energy is
rapidly converted into potential, kinetic and gradient energy of the waterfall field
and, if the motion of the inflaton field is taken into account, also
into gradient energy of the inflaton field \cite{gb2}. 
In order to determine the various contributions to the energy density,
one has to take the couplings of all modes of the inflaton and waterfall
field into account \cite{Kofman:1997yn}.
For couplings $\mathcal{O}(1)$, all these energy densities are
$\mathcal{O}(v^4)$, where $v$ is the symmetry breaking vacuum
expectation value. Hence, the equation of state changes and inflation ends.
In particular the model of Ref.~\cite{gb2} is very similar to case of 
D-term inflation considered in this paper, except for one crucial
difference. In Ref.~\cite{gb2}, the self-coupling of the
waterfall field and the coupling of the waterfall field to the
inflaton have equal strength, $\lambda/g = 2$ (cf.~\eqref{eq_V2}). 
As a consequence, in Ref.~\cite{gb2} it is found
that waterfall field and inflaton field both rapidly approach the
ground state, performing together coherent oscillations. 
In this paper, on the contrary, $\lambda/g \simeq 7\times 10^{-4}$.
Hence, the inflaton motion is much less affected by the growth of
the waterfall field which in turn approches an inflaton-dependent 
minimum $s^2_{\rm min}(\varphi)$.

In order to estimate the effect of self-interaction of the waterfall
field we first determine the decoherence time $t_{\rm dec}$ where
the waterfall field becomes classical. This occurs if
the product of $s_k$ and the
canonically conjugate momentum $\pi_{sk}$ is much larger than $\hbar/2$,
the minimal value for an oscillating mode \cite{Guth:1985ya}. Demanding that
$|s_k(t_{\rm dec})\pi_{sk}(t_{\rm dec})| \equiv R_{\rm dec} \gg \hbar \equiv 1$, one finds
for soft modes $k < k_b$,
\begin{equation}
t_{\rm dec} \sim \frac{1}{D}\left[\tfrac{3}{4}\ln(2 R_{\rm
    dec})\right]^{2/3} \ .
\end{equation}
For the parameter values above and $R_{\rm dec} = 100$, one obtains $t_{\rm
  dec} \simeq 1.7/H_c$. As expected, one has for times $t \geq t_{\rm dec}$,
\begin{equation}
\frac{\dot s}{H_c} > \frac{H_c}{2\pi} \ ,
\end{equation}
i.e.\ the classical growth dominates over the
quantum growth. In particular, for the parameter example above, we find 
$\dot s/H_c \simeq 3\times 10^{-5}$ and $H_c/(2 \pi) \simeq 6 \times 10^{-6}$ at $t=t_{\textrm{dec}}$.

\begin{figure}[t]
 \includegraphics[width = 0.47\textwidth]{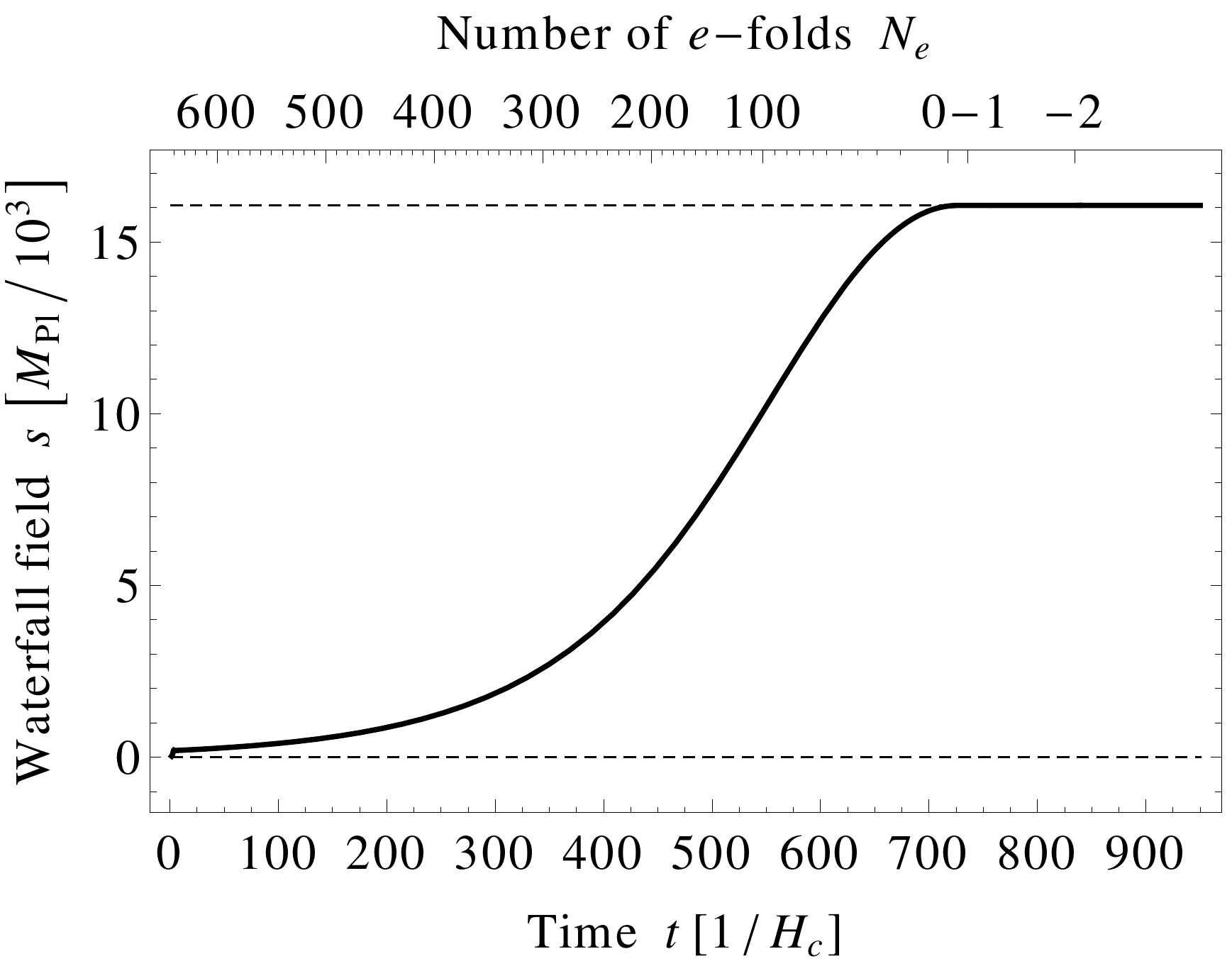}\hspace{0.8cm}\includegraphics[width = 0.47\textwidth]{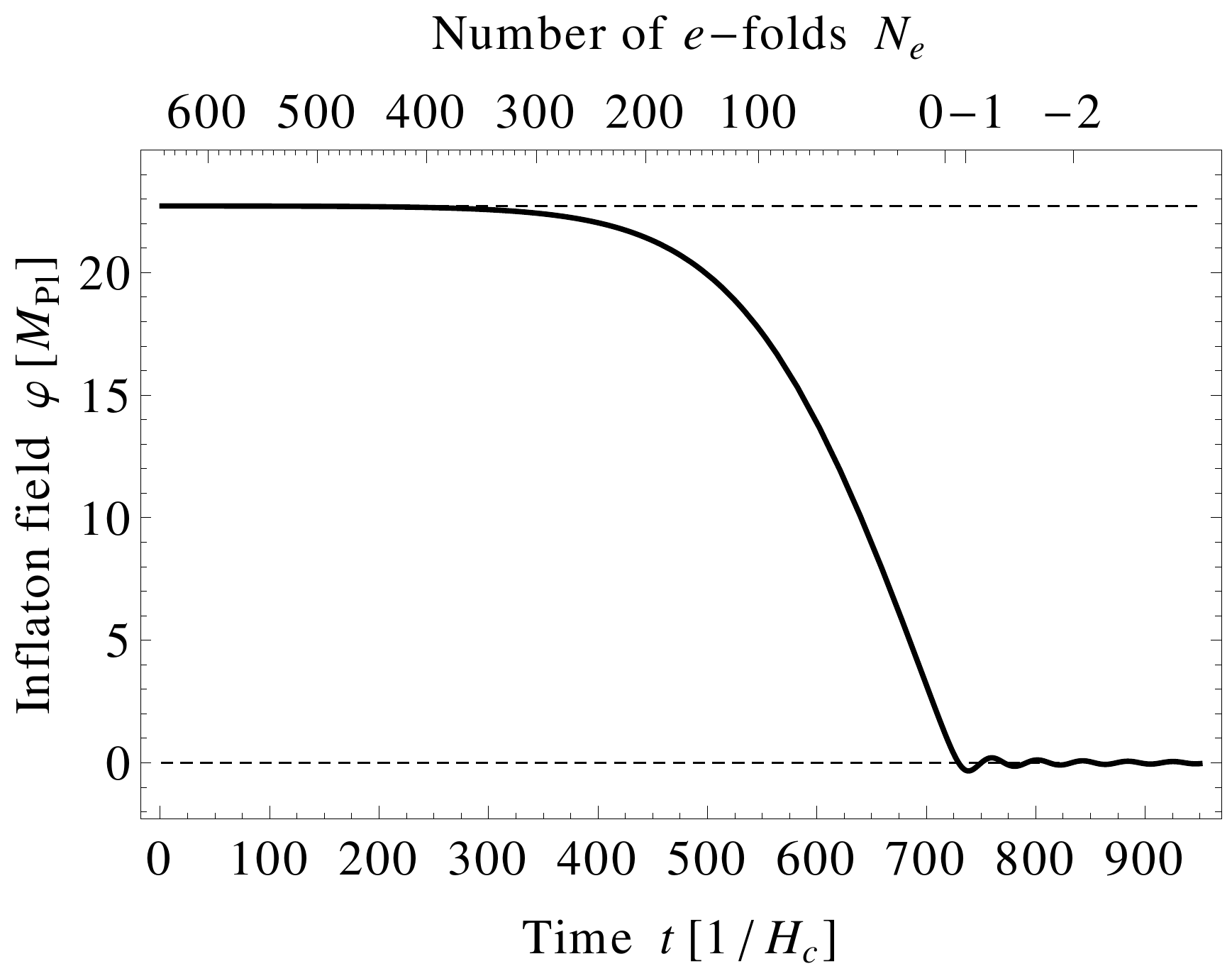}
 \caption{Classical field evolution of the waterfall (left) and inflaton (right) field. }
 \label{fig_classical}
\end{figure}

At the decoherence time $t_\text{dec}$, the classical waterfall field can be approximated
by $s(t_{\rm dec}) = \langle s^2(t_{\rm dec}) \rangle^{1/2}$. Its initial value
and its velocity can then be read off from Fig.~\ref{fig_tp}a. With
these initial values one can solve the classical field equations
\begin{align}
\ddot \vp + 3H\dot \vp + \frac{1}{2} \lambda^2 s^2 \vp &= 0 \ ,  \label{dg_phi}\\
\ddot s + 3H\dot s - \left(g^2\xi - \frac{\lambda^2}{2}\vp^2\right)s +
\frac{g^2}{2}s^3 &= 0 \ .
\label{dg_s}
\end{align}
The result is shown in Fig.~\ref{fig_tp}b.
In contrast to our estimate for the variance $\left<s^2(t)\right>$ in
Fig.~\ref{fig_tp}a, the solution of Eq.~\eqref{dg_s} is affected by
the quartic self-interaction of the waterfall field,
which is reflected in $s(t)$ growing significantly more slowly than
$\left<s^2(t)\right>$ after $t=t_{\textrm{dec}}$.
At $t\simeq 3.4/H_c$, the waterfall field reaches for the first time
the inflaton-dependent minimum.
Initially, it overshoots the minimum; but after only one oscillation
it basically becomes stabilized at $s(t) = s_{\textrm{min}}\left(\varphi(t)\right)$.
The tachyonic growth of the waterfall field modes leads to kinetic and
gradient energy. Since the dominant momenta are larger than the
Hubble scale, this process can be expected to proceed analogous to 
tachyonic preheating, i.e.\ the corresponding energy densities are
$\mathcal{O}(s_{\rm min}^4)$. Since at $t^{\rm loc}_{\rm sp}$ the
expectation of the waterfall field is $\langle s^2(t_{\rm sp}^{\rm
  loc}) \rangle^{1/2} \sim 10^{-2} s_0$, one finds that these energy
densities are suppressed by 8 orders of magnitude compared to the
dominant vacuum energy $V_0 = g^2\xi^2/2 = g^2 s_0^4/8$.
Hence, contrary to standard tachyonic preheating, inflation continues
despite the tachyonic growth of the waterfall field.  

At later times, the waterfall field then simply follows the position of 
the local minimum adiabatically.
The global minimum is reached much later around $t \simeq 720/H_c$.
The time evolution of the inflaton-waterfall system is
depicted in Fig.~\ref{fig_classical}. After the waterfall field has reached its global
minimum, the inflaton performs small oscillations around its
minimum, leading to standard reheating.

\section{An emerging regime of chaotic inflation \label{sec_4}}

\begin{figure}
\begin{center}
 \includegraphics[width = 0.65\textwidth]{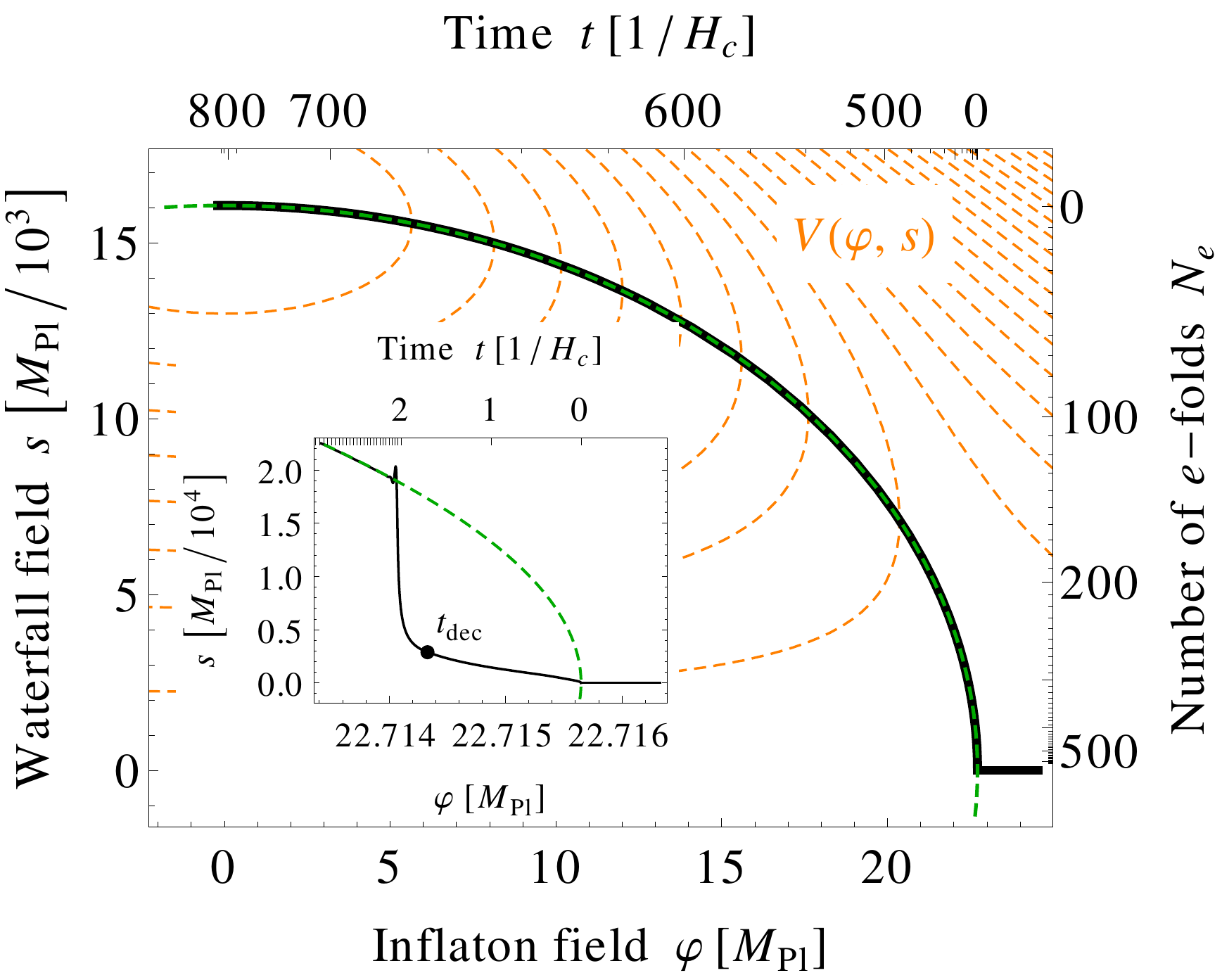}\\
\vspace{1cm}
\includegraphics[width = 0.65\textwidth]{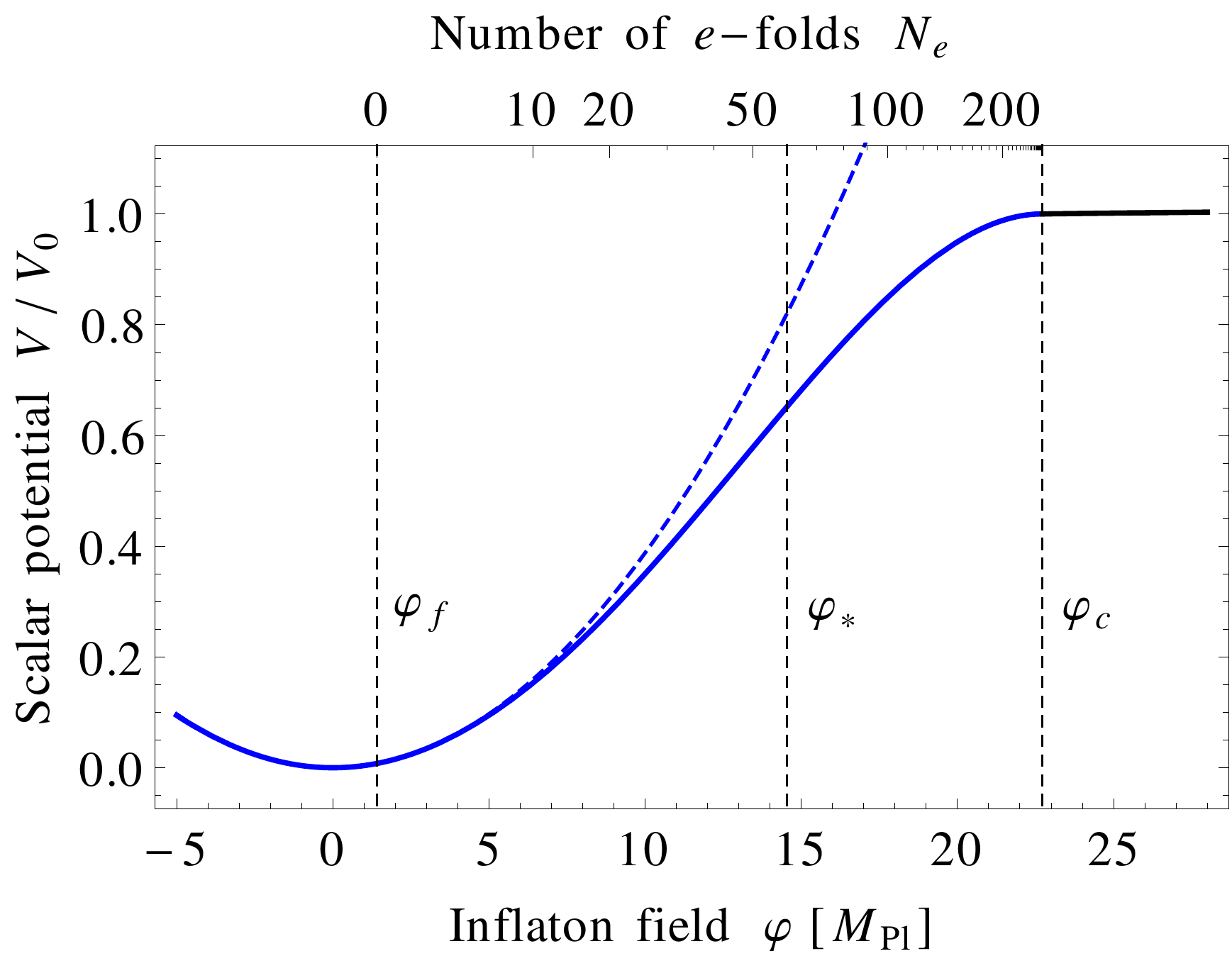}
\end{center}
\caption{Effective scalar potential during the second stage of inflation. \textbf{Upper (a):} Classical trajectory in two-dimensional field space. Dashed green: analytical solution~\eqref{s2min}, black: numerical solution from Eqs.\ \eqref{dg_phi} and \eqref{dg_s}.
\textbf{Lower (b):} Effective scalar potential for the inflaton field after integrating out the waterfall field. Solid black: Tree-level plus 1-loop potential, solid blue: substituting the waterfall field by its local minimum~\eqref{s2min}, dashed blue: quadratic approximation. 
}
\label{fig_potential}
\end{figure}

After a short initial period of tachyonic preheating, the waterfall
field tracks the inflaton-dependent minimum $s_{\rm min}(t)$,
cf.\ Eq.~\eqref{s2min}, and both, inflaton and
waterfall field perform a coupled slow-roll motion towards the ground
state. The corresponding trajectory in field space is shown in
Fig.~\ref{fig_potential}a.
Along almost the entire trajectory, we find
\begin{align}
\frac{\partial^2 V}{\partial s^2}\  &\gg\  \frac{\partial^2
  V}{\partial\vp\partial s}\ \gg\ \frac{\partial^2 V}{\partial \vp^2}
\ ,\\
\frac{\partial^2 V}{\partial s^2}\  &\gg\  H_c^2\
\gg\ \frac{\partial^2 V}{\partial \vp^2} \ .
\end{align}
Hence, one essentially has a one-field model of inflation, and the
curvature perturbations are dominated by the quantum fluctuations of
the field $\vp$.

The effective inflaton potential is obtained by inserting the instantaneous minimum of the waterfall
field \eqref{s2min}\footnote{Note that $S_-$ plays the role of the
  stabilizer field in supersymmetric chaotic inflation, for a which a
  quartic term in the K\"ahler potential is generated by integrating
  out the massive vector field, which is consistent near the global
  minimum~\cite{Wieck:2014xxa}.} 
in the potential \eqref{eq_V2}, 
\begin{equation}
 V(\vp,s_{\rm min}(\vp)) =  \frac{1}{2}\lambda^2\xi\,\vp^2 \left(1 - \frac{1}{2}\frac{\vp^2}{\vp_c^2}\right) \,.
\label{Vinf}
\end{equation}
Here we have neglected the loop-suppressed radiative corrections due to the interaction between the inflaton and the waterfall fields.
Note that ratio the $\lambda/g$, which determines the position of $\varphi_c$ in Fig.~\ref{fig_potential}b, also determines the ratio of the semi-axes of the ellipse shown in Fig.~\ref{fig_potential}a, cf.\ Eq.~\eqref{s2min}. A small value of this ratio and hence transplanckian value for $\varphi_c$ is crucial for our scenario: it ensures sufficient slow-roll inflation along the inflaton direction \textit{after} the critical point.
Remarkably enough, for small field values, $\vp \ll \vp_c = g/ \lambda \,\sqrt{2\xi}$,
Eq.~\eqref{Vinf} closely resembles the potential of the simplest example of chaotic inflation,
$V(\vp) \simeq  m_I^2\vp^2/2$, with $m_I^2 = \lambda^2\xi$, cf.\ Fig.~\ref{fig_potential}b.
We shall now calculate our predictions for the inflationary observables based
on Eq.~\eqref{Vinf} and demonstrate how these reduce to the well-known expressions
for chaotic inflation in the limit $\lambda/g \ll 1$, i.e.\ for large
critical inflaton values, $\varphi_c \gg 1$.

The scalar spectral amplitude $A_s$, the scalar spectral index $n_s$ and
the tensor-to-scalar ratio $r$ can all be expressed in terms of the
slow-roll parameters $\epsilon$ and $\eta$,
\begin{equation}
 \begin{split}
  \epsilon &= \frac{1}{2} \left(\frac{V'}{V} \right)^2 =
\frac{8 \left(\varphi_c^2 -
    \varphi^2\right)^2}{\varphi^2\left(2\varphi_c^2 - \varphi^2\right)^2}
\simeq \frac{2}{\varphi ^2} - \frac{2}{\varphi_c^2}
\,, \\  \eta &=  \frac{V''}{V} = 
\frac{2}{\varphi ^2} - \frac{10}{2 \varphi_c^2 - \varphi^2} \simeq 
\frac{2}{\varphi ^2} - \frac{5}{\varphi_c^2} \,.
 \end{split}
\end{equation}

Inflation ends when the slow-roll parameters
approach one, i.e.\ at $\varphi = \varphi_f \equiv
\textrm{max}\left\{\varphi_\epsilon,\varphi_\eta\right\}$,
where $\varphi_\epsilon$ and $\varphi_\eta$ are defined such that
$\epsilon(\varphi_\epsilon) = 1$ and $\eta(\varphi_\eta) = 1$,
respectively. We have
\begin{align}
\varphi_\epsilon^2 \simeq 2 - \frac{4}{\varphi_c^2} \,, \quad 
\varphi_\eta^2 = 6 + \varphi_c^2 \left[1-\left(1+\frac{8}{\varphi_c^2}
+ \frac{36}{\varphi_c^4}\right)^{1/2}\right]
\simeq 2 - \frac{10}{\varphi_c^2} \,,
\end{align}
so that $\varphi_\epsilon$ is always larger than $\varphi_\eta$ and
hence $\varphi_f \equiv \varphi_\epsilon$.
Solving the slow-roll equation yields the inflaton value $\vp(N_e)$ as a function of
$N_e$, the number of $e$-folds before the end of inflation,
\begin{equation}
 \int_{\vp_f}^{\vp(N_e)} \frac{V}{V'} \,d \varphi = N_e \quad \rightarrow \quad
 \vp^2(N_e) = \varphi_c^2 \left[1 - W_0\left(\Delta \,
 e^\Delta \,e^{-8N_e/\varphi_c^2}\right)\right] \,, 
\end{equation}
Here,  $\Delta \equiv 1 - \varphi_f^2/\varphi_c^2$ and $W_0$ denotes the principal branch of the Lambert $W$ function or product logarithm,
which can take values $W_0\geq -1$ and which satisfies $x = W_0(x)\,e^{W_0(x)}$,
so that $W_0\left(x \,e^x\right) = x$.
For large $\varphi_c$, we obtain the familiar expression from chaotic inflation
to leading order,
\begin{align}
\varphi^2(N_e) \simeq \left(4N_e + 2\right) -\frac{4}{\varphi_c^2}
\left(N^2_e + N_e + 1\right) \,.
\label{eq:phiN}
\end{align}
The inflationary observables at $\varphi_*$,  $N_* \simeq 60$ $e$-folds before the end of inflation, are given by
\begin{equation}
A_s \simeq 
\frac{\lambda ^4 \varphi_* ^4 \varphi_c^2}{192 \pi ^2 g^2}\left(1 +
\frac{1}{2}\frac{\varphi_*^2}{\varphi_c^2}\right) \,, \quad  
n_s \simeq 1-\frac{8}{\varphi_* ^2}\left(1 - \frac{1}{4}\frac{\varphi_*^2}{\varphi_c^2}\right) \,,
\quad r \simeq \frac{32}{\varphi_* ^2}\left(1 - \frac{\varphi_*^2}{\varphi_c^2}\right) \,. 
\label{eq:obs}
\end{equation}
Again, the leading terms are the usual expressions known from chaotic inflation.
In the limit of chaotic inflation, the effective inflaton mass is fixed by
the measured amplitude of the power spectrum~\cite{Ade:2013uln},
\begin{equation}\label{norm}
m_I^2 \simeq \frac{96 \,\pi^2}{\varphi_*^4}  A_s^\text{obs} \simeq \left( 1.4 \times 10^{13}~\text{GeV} \right)^2\,, \quad A_s^\text{obs} \simeq 2.2 \times 10^{-9}  \,,
\end{equation}
where we employed the expression for $A_s$ in Eq.~\eqref{eq:obs} as well
as the relation $\varphi_c = g/\lambda\sqrt{2\xi}$.
This condition can also be used to derive an upper bound on the coupling $\lambda$.
For larger values of $\lambda$, the separation between $\varphi_c$ and $\varphi_*$
becomes increasingly smaller, until $\varphi_c$ eventually reaches a minimal value
$\varphi_c^{\textrm{min}}$ and $\varphi_* \simeq \varphi_c = \varphi_c^{\textrm{min}}$.
The difference between our result for $\varphi_*$ in Eq.~\eqref{eq:phiN}
and $\varphi_c$ vanishes for $\varphi_c \rightarrow \varphi_c^{\textrm{min}}
\simeq \left(2N_* + 1\right)^{1/2} \simeq 11$.
In order to realize $N_*$ $e$-folds of inflation after the critical point,
we must therefore require 
\begin{equation}
 \lambda \lesssim \left(\frac{192 \pi^2 g^2}{\left(\varphi_c^{\textrm{min}}\right)^6} A_s^\text{obs}\right)^{1/4} \simeq 1\times 10^{-3}\,, \quad  \sqrt{\xi} 
 \gtrsim \left(\frac{12 \pi^2}{g^2 \left(\varphi_c^{\textrm{min}}\right)^2} A_s^\text{obs}\right)^{1/4}  \simeq
 2 \times 10^{16}\,\textrm{GeV} \,,
\label{bound1}
\end{equation}
where we have used Eqs.~\eqref{eq:obs} and \eqref{norm} and with
$g^2$ and $N_*$ being set to $1/2$ and $60$, respectively. On the other hand, for very
small values of $\lambda$ the discussion of this paper is no longer
applicable. A rough lower bound on $\lambda$ can be obtained by requiring that the
local minimum of the waterfall field at $t_{\text dec}$ lies outside
the quantum uncertainty of the false vacuum at the origin,
$s_\text{min}(t_{\text dec}) > H_c/(2 \pi)$,
\begin{equation}
 \lambda \gtrsim 1 \times 10^{-4} \,, \quad \sqrt{\xi} \lesssim 2\times 10^{17}\,\textrm{GeV} \,.
\label{bound2}
\end{equation}
Approaching these values, the time scales of tachyonic preheating are stretched to ${\cal O}(100)$ Hubble times. A consistent description of the phase transition in this regime requires further investigation. In any case, the bounds~\eqref{bound1} and \eqref{bound2} show that a
regime of chaotic inflation can indeed be obtained. The allowed ranges for $\lambda$ and $\xi$ 
are basically found to cover one order of magnitude, respectively.
Remarkably enough, this regime
coincides with an initial GUT-scale energy density!

From Fig.~\ref{fig_potential}b we note that, for the parameter point discussed in this paper ($g^2 = 1/2$, $\lambda = 5 \times 10^{-4}$), there is a sizable deviation from the quadratic potential at $\varphi = \varphi_*$. Using the above expressions based on the exact potential, we find
\begin{equation}
 \varphi_* = 14.5 \, M_{\textrm{Pl}} \,, \quad n_s = 0.963 \,, \quad r = 0.083 \,.
\end{equation}
Here, the value of $\sqrt{\xi} = 2.8 \times 10^{16}$~GeV has been fixed to obtain the correct amplitude, cf.\ Eq.~\eqref{norm}.
These results can be compared with the predictions in the purely quadratic approximation, which
corresponds to the limit $\varphi_c\rightarrow\infty$ for fixed effective inflaton mass $m_I$.
In this approximation, we find $\varphi_* \simeq 15.5\,M_\textrm{Pl}$,
$A_s \simeq 2.0 \times 10^{-9}$, $n_s \simeq 0.967$ and $r \simeq 0.133$.
Hence the quadratic approximation provides quite good results in the parameter range under study. Our explicit calculation based on the full potential however shows that in particular the tensor-to-scalar ratio ends up being somewhat smaller than in the quadratic approximation. This is evident from the expression for $r$ in Eq.~\eqref{eq:obs}.

\section{Cosmic strings \label{sec_5}}

In the usual formulation of D-term hybrid inflation, cosmic strings produced during the phase transition at the end of inflation impose serious constraints. For typical GUT-scale values,  $\sqrt{\xi} \gtrsim 10^{16}$~GeV, the model is ruled out by the non-observation of cosmic strings in the CMB~\cite{Ade:2013xla}. The new inflationary regime identified in Section \ref{sec_4} radically changes this picture. Cosmic strings are formed during tachyonic preheating with an average distance of
\begin{equation}
 a \, k_*^{-1} \sim  p_b^{-1}(t_\text{sp}^\text{loc}) =   {\cal O} \left( \frac{1}{H_c} \right) \,,
\end{equation}
which implies that after $\sim 60$ $e$-folds of inflation they have been 'inflated away' to unobservably large scales. In particular for the parameter example discussed in Sections~\ref{sec_3} and \ref{sec_4}, many $e$-folds occur between string formation at the critical point $\varphi_c$ and the onset of the last 60 $e$-folds at $\varphi_*$. In this case, cosmic strings  cannot produce any observable effects.\footnote{For accordingly tuned parameters (relatively large $\lambda$ close to the bound in Eq.~\eqref{bound1}), one could obtain $\varphi_* \simeq \varphi_c$. In this case, cosmic strings would be just on the boundary of our observable universe. One might speculate if the anomalies observed in the CMB at very low multipoles could be consistent with such a scenario.}

Extending the parameter space beyond Eq.~\eqref{bound1} to larger values of $\lambda$, $60$ $e$-folds of inflation no longer `fit' into the time interval after the phase transition. Instead, one might picture a situation where the final $N_*$ $e$-folds of inflation are split between the Coleman-Weinberg regime before the phase transition and the chaotic regime after the phase transition and where cosmic string production occurs during inflation. This is similar to the situation considered in \cite{Linde:2013aya} where cosmic strings are produced during a dominantly chaotic phase of inflation. In this case, a sufficient amount of inflation after the phase transition relaxes the cosmic string problem as cosmic string signatures would be absent on small scales. However, observable effects in the CMB, the gravitational wave spectrum or the large scale galaxy distribution may still be present and require a more detailed analysis, cf.\ Refs.~\cite{Kofman:1986wm, Vishniac:1986sk}.

\section{Conclusion and Outlook \label{sec_6}}

Standard hybrid inflation ends in a waterfall transition at a critical
value $\vp_c$ of the inflaton field. Due to the rapid growth of
quantum fluctuations, the waterfall field reaches a classical regime within
a few Hubble times where, due to backreaction effects, it settles at
an inflaton-dependent, instantaneous minimum $s_{\rm min}(t)$. This
initial stage of tachyonic preheating is followed by a coupled slow-roll 
motion of the inflaton-waterfall system towards the ground state. 
For sufficiently small values of the inflaton coupling to the waterfall 
field, this post-critical period of inflation can last longer than 60
$e$-folds. As the waterfall field approaches its ground state, it
generates an effective mass term for the inflaton field, leading
to the standard predictions of chaotic inflation.

This emerging regime of chaotic inflation has several remarkable
features. First, it starts from a constant energy density close to the
scale of grand unification. If a sizeable amplitude of primordial gravitational waves should be indeed confirmed by further measurements in the near future, pointing to inflation at the GUT scale, this setup would provide a quite unique possibility of explaining the appearance of this scale at transplanckian field values.
Second, since the relevant phase of inflation
takes place after symmetry breaking, there is no cosmic string problem.
Finally, compared to the simplest example of chaotic inflation based on a
quadratic potential, a somewhat smaller tensor-to-scalar ratio is predicted whereas
the scalar spectral index is identical. 

The crucial parameter which governs the existence of this new regime is the ratio of the superpotential and the gauge coupling, $\lambda/g \ll 1$. The smaller this value, the more the time-scales relevant for the phase transition are stretched out. This implies that the inflaton velocity and the Hubble expansion play a crucial role in the tachyonic preheating process, while simultaneously enabling a phase of single-field slow-roll inflation in the inflaton direction after passing the critical value $\varphi_c$. Note that this is not possible for F-term hybrid inflation, where the self-interaction of the waterfall field and its coupling to the inflaton field are determined by the same coupling constant.

\medskip \medskip


\noindent \textbf{Acknowledgments} \medskip

\noindent We are grateful to Juan Garc\'ia-Bellido, Arthur Hebecker, Andrei Linde and Wei Xue for very helpful discussions and comments. 
In particular, V.D.\ and K.S.\ thank Tsutomu Yanagida for many inspiring discussions, which triggered some of the ideas for this project.
In this context V.D.\  gratefully acknowledges the hospitality of  Kavli IPMU during the initial stage of this project.
This work has been supported in part by the German Science Foundation (DFG)
within the Collaborative Research Center 676 ``Particles, Strings and the Early Universe'' (W.B.), by the European Union FP7-ITN INVISIBLES (Marie Curie Action PITAN-GA-2011-289442-INVISIBLES) (V.D.) and by the World Premier International Research Center Initiative (WPI) of the Ministry of Education, Culture,
Sports, Science and Technology (MEXT) of Japan (K.S.).

\end{document}